\documentclass[12pt]{article}
\usepackage{wrapfig}
\usepackage{epsfig}
\usepackage{hyperref}
\usepackage{amssymb}
\usepackage{amsmath,mathtools}
\usepackage{amsfonts}
\usepackage[mathscr]{euscript}
\usepackage{latexsym}
\usepackage{wasysym}
\usepackage{multirow}
\usepackage{fixmath}
\usepackage{color}
\usepackage{stackrel}
\usepackage{txfonts}                                                            \usepackage{centernot}
\usepackage{cancel}
\usepackage{bbm}
\usepackage[low-sup]{subdepth}
\usepackage[symbol]{footmisc}
\usepackage{stackengine}

\newcommand{\bea}{\begin{eqnarray}}
\newcommand{\eea}{\end{eqnarray}}
\newcommand{\bite}{\begin{itemize}}
\newcommand{\eite}{\end{itemize}}

\newcommand{\phibar}{\hskip1.0mm{\bar{\hskip-1.0mm\phi}}}

\date{}

\begin{document}
\title{
\vspace{-2.0cm} 
\vspace{0.5cm}
\centering{\Large \bf The QCD axion beyond the classical level: A lattice study}}

\author{Y.~Nakamura$^a$ and G.~Schierholz$^b$\\[1em] 
$^a$ RIKEN Advanced Institute for Computational Science,\\ Kobe, Hyogo 650-0047, Japan\\[0.15em] 
$^b$ Deutsches Elektronen-Synchrotron DESY,\\ 22603 Hamburg, Germany}

\maketitle
\vspace*{-0.5cm}

\begin{abstract}
The axion is a hypothetical elementary particle postulated by the Peccei-Quinn theory to resolve the strong CP problem in QCD. If axions exist and have low mass, they are a candidate for dark matter as well. So far our knowledge of the properties of the QCD axion rests on semi-classical arguments and effective theory. In this work we perform, for the first time, a fully dynamical investigation of the Peccei-Quinn theory, focussing on the axion mass, by simulating the theory on the lattice. The results of the simulation are found to be in conflict with present axion phenomenology. 
\end{abstract}

\newpage
\section{The strong CP problem and axion}

Quantum chromodynamics (QCD) decribes the strong interactions remarkably well down to the smallest scales probed so far. Yet it faces a problem. The theory allows for a CP-violating term $S_\theta$ in the action,
\begin{equation}
S = S_{\rm QCD}+S_\theta\,,
\label{CPact}
\end{equation}
the so-called $\theta$ term. In Euclidean space-time $S_\theta$ reads
\begin{equation}
S_\theta = i\, \theta\, Q\,, \;\; Q = \int\! d^4x\, q(x)\, \in\, \mathbb{Z}\,,
\label{thetaterm}
\end{equation}
where $Q$ is the topological charge with charge density
\begin{equation}
q(x) = - \frac{1}{64\pi^2} \, \epsilon_{\mu\nu\rho\sigma}\,F_{\mu\nu}^a(x)\,F_{\rho\sigma}^a(x)\,.
\label{topcharge}
\end{equation}
In this formulation $\theta$ enters as an arbitrary phase with values $-\pi < \theta \leq \pi$, referred to as the vacuum angle. The problem is that no CP violation has been observed in the strong interactions. A nonvanishing value of $\theta$ would result in an electric dipole moment $d_n$ of the neutron. Current experimental limits on $|d_n|$~\cite{Baker}, paired with lattice calculations~\cite{Guo:2015tla}, lead to the upper bound $\displaystyle |\theta| \lesssim 7.4 \times 10^{-11}$. This anomalously small number is referred to as the strong CP problem, which is one of the most intriguing problems in particle physics. 

In the Peccei-Quinn theory~\cite{Peccei:1977hh} the CP violating action $S_\theta$ is augmented by the axion interaction 
\begin{equation}
S_\theta \rightarrow S_\theta + S_{\rm Axion} = \int \! d^4x \, \left[\frac{1}{2} \big(\partial_\mu \phi_a(x)\big)^2 + i \left(\theta + \frac{\phi_a(x)}{f_a}\right) \, q(x) \right]\,,
\label{axact1}
\end{equation}
where $\phi_a(x)$ is the axion field and $f_a$ is the axion decay constant. The basic idea is to raise the vacuum angle $\theta$ to a dynamical variable. It is then expected that QCD induces an effective potential for $\theta +\phi_a/f_a$, $U_{\rm eff}(\theta +\phi_a/f_a)$, having a stationary point at $\theta +\phi_a/f_a=0$, thus restoring CP symmetry.

In the following we will treat the axion field as a dynamical degree of freedom, whose purpose is to solve the strong CP problem, whether it arises from the spontaneously broken Peccei-Quinn symmetry or from a more fundamental theory, and focus on QCD interactions. The strong CP problem is a problem of QCD, which calls for a solution on the hadronic scale. Any higher interactions can be considered to be integrated out.

\section{Preliminaries}

We are primarily interested in the axion mass, which is the keystone of present-day axion phenomenology. Throughout this work we will consider the case $\theta = 0$ only. This leaves us with the action
\begin{equation}
S = S_{\rm QCD} + S_{\rm Axion}\,, \;\; S_{\rm Axion} = \int \! d^4x \, \left[\frac{1}{2} \big(\partial_\mu \phi_a(x)\big)^2 + i\,  \frac{\phi_a(x)}{f_a} \, q(x) \right]\,,
\label{axact2}
\end{equation}
which now includes quarks and gluons. This is one of several axion actions, which are equivalent at the QCD level. It is referred to as the Kim-Shifman-Vainshtein-Zakharov (KSVZ) parameterization of the action in the literature~\cite{Kim:1986ax}. For the most general action see the Appendix.


In classical approximation $\phi_a(x)$ is treated as an external source, $\phi_a(x) \rightarrow \phibar_a$, with
\begin{equation}
\phibar_a = \frac{1}{V}\int d^4x\,\phi_a(x)\,,
\end{equation}
where $V$ is the space-time volume. The topological charge $Q$ is found to follow a Gaussian distribution~\cite{DelDebbio:2004ns}. The partition function can then be written
\begin{equation}
Z_{\rm class} = \frac{1}{\sqrt{2\pi\langle Q^2\rangle}}\int dQ\, d\phibar_a\, \exp\left\{-Q^2/2\langle Q^2\rangle - i\,(\phibar_a/f_a)\,Q - (\mu_a^2/2)\,\phibar_a^2\,V\right\}\,,
\label{part}
\end{equation}
where we allowed for a hypothetical mass term. If QCD effects are neglected, the axion would be massless, and any value of $\phi_a$ would be equally acceptable from the energetic point of view. In particular, the theory would be invariant under the shift $\phi_a \rightarrow \phi_a + \delta$. This is no longer guaranteed if QCD effects are taken into account. From (\ref{part}) we derive
\begin{equation}
\langle \phibar_a^2 \rangle = \frac{1}{(\chi_t/f_a^2+ \mu_a^2)\, V}\,, \quad \chi_t = \frac{\langle Q^2\rangle}{V} \,,
\label{fsize}
\end{equation}
where $\chi_t$ is the topological susceptibility. It follows that only a limited range of $\phi_a$ values is allowed, which will shrink to zero as the volume $V$ is taken to infinity.  
 
The partition function (\ref{part}) generates a classical potential $U_{\rm class}(\phibar_a)$ for the axion field,
\begin{equation}
\frac{1}{\sqrt{2\pi\langle Q^2\rangle}}\int dQ\, \exp\left\{-Q^2/2\langle Q^2\rangle - i\,(\phibar_a/f_a)\,Q\right\} = \exp\left\{-V U_{\rm class}(\phibar_a)\right\}\,.
\label{partphi}
\end{equation}
The result is 
\begin{equation}
U_{\rm class}(\phibar_a) = \frac{\chi_t}{2f_a^2}\, \phibar_a^2 \,.
\label{classpot}
\end{equation}
Expanding the potential (\ref{classpot}) at the minimum, $\phibar_a = 0$, gives the axion a mass~\cite{Peccei:2006as},
\begin{equation}
m_a^2 = \frac{\partial^2}{\partial \phibar_a^2}\, U_{\rm class}(\phibar_a)\,\big|_{\,\phibar_a=0} = \frac{\chi_t}{f_a^2}\,.
\label{axmass}
\end{equation}
The right-hand side of (\ref{axmass}) can be viewed as a sort of vacuum energy. Following Weinberg~\cite{Weinberg:1977ma} and Wilczek~\cite{Wilczek:1977pj}, the QCD axion is commonly interpreted as the pseudo-Goldstone boson of the Peccei-Quinn symmetry broken by the anomalous coupling to the topological charge. The expression (\ref{axmass}) for the axion mass is widely used in phenomenological applications, notably in models of dark matter. If correct, the axion could be made practically invisible by taking $f_a$ to infinity.

\begin{figure}[t]
\begin{center}
\epsfig{file=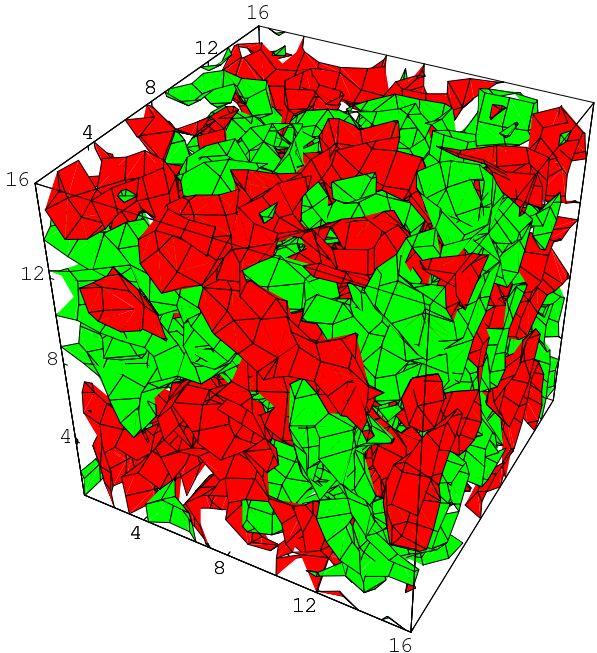,width=8cm,clip=}
\end{center}
\vspace*{-0.5cm}
\caption{A typical snapshot of the QCD vacuum. The picture shows a single time slice of a $16^4$ lattice. The vacuum is composed of isosurfaces of positive (red) and negative (green) topological charge density, which are infinitely thin in the co-direction and form a multi-layered compound that covers all space-time.}
\label{vac}
\end{figure}

The classical axion potential (\ref{classpot}) is a special case of the effective potential~\cite{Fukuda:1974ey,ORaifeartaigh:1986axd}, which is given by 
\begin{equation}
V\,U_{\rm eff}(\phibar_a)= - \log P(\phibar_a) + c \,,
\label{effpotphi}
\end{equation}
where $P(\phibar_a)$ denotes the probability of finding an axion field of value $\phibar_a$ in the vacuum. In Fig.~\ref{vac} we show a snapshot of the toplogical charge density $q(x)$ of the QCD vacuum~\cite{Ilgenfritz:2005hh,Ilgenfritz:2007xu}, which the axion field $\phi_a(x)$ is exposed to. This was made possible by employing a fermionic definition of $q(x)$ using overlap fermions~\cite{Hasenfratz:1998ri}. It turned out that the QCD vacuum decomposes into isosurfaces of alternating topological charge density that swamp any underlying semi-classical topological structure. Responsible for that are the low-lying, continuous eigenmodes of the Dirac operator, rather than the Atiyah-Singer zero modes. This result can hardly be reconciled with the picture of a dilute gas of instantons. On the periodic lattice the axion Lagrangian takes its minimum at
\begin{equation}
\phi_a(x) = \frac{2i}{f_a}\, \int \!d^4y\, \Delta^{-1}(x-y)\, q(y) \equiv \frac{2i}{f_a} \Delta^{-1}q(x)\,,
\end{equation}
where $\Delta^{-1}$ is the inverse Laplacian, which indicates that the axion will closely follow the quantum fluctuations of the topological charge density. This will inevitably lead to significant corrections to the classical potential (\ref{classpot}) and axion mass (\ref{axmass}).  
 
In Euclidean quantum field theory the axion mass $m_a$ is given by the large-time $t$ decay of the axion correlation function, 
\begin{equation}
\int\! d^3\vec{x}\,\left\langle \phi_a(\vec{x},t)\, \pi(0)\right\rangle \simeq A\, {\rm e}^{-m_a t} \,,
\label{phipi}
\end{equation}
where $\pi$ is a pseudoscalar source. The underlying path integral does not change under the shift of integration variables $\phi_a \rightarrow \phi_a + \epsilon_a$, nor does the shift alter the integration measure. Expanding the path integral to first order in $\epsilon_a$, we obtain the equation of motion
\begin{equation}
\frac{\partial^2}{\partial t^2} \int\! d^3\vec{x}\,\left\langle \phi_a(\vec{x},t)\, \pi(0)\right\rangle = \frac{i}{f_a}\,\int\! d^3\vec{x}\,\left\langle q(\vec{x},t)\, \pi(0)\right\rangle\,, \;\; t>0 \,.
\label{eqmot}
\end{equation}
For the pseudoscalar source $\pi$ we can take any field that couples to the axion field $\phi_a$. Taking $\pi=q$, and assuming $1/f_a \neq 0$, we arrive at the expression for the axion mass 
\begin{equation}
m_a = - \lim_{t\rightarrow\infty} \frac{1}{t} \log \int\! d^3\vec{x}\, \langle q(\vec{x},t)\,q(0) \rangle  \,. 
\label{ma1}
\end{equation}
For $\pi = \phi_a$ the equation of motion (\ref{eqmot}) allows for an exactly massless particle, in addition to (\ref{ma1}), that does not couple to $q(x)$. Making use of the axial anomaly, the correlator in (\ref{ma1}) can be expressed entirely in terms of quark fields, 
\begin{equation}
m_a =  - \lim_{t\rightarrow\infty} \frac{1}{t} \log \int\! d^3\vec{x}\, \langle P(\vec{x},t)\,P(0) \rangle_{disc} \,, 
\label{ma2}
\end{equation}
where $P$ is the pseudoscalar density, $P=\big(\bar{u}\gamma_5u + \bar{d}\gamma_5d + \cdots\big)$, and {\it disc} stands for fermion-line disconnected. This suggests that the axion will strongly mix with the pseudoscalar meson sector.

In the weak coupling limit, $1/f_a \rightarrow 0$, we expect the axion mass $m_a$ to be driven almost exclusively by QCD interactions, with little feedback of $\phi_a(x)$ on $q(x)$ and $P(x)$, and thus become largely independent of $f_a$. To substantiate this claim, consider the correlator $\int\! d^3\vec{x}\, \langle q(\vec{x},t)\,q(0) \rangle$ in (\ref{ma1}). In this case the axion field $\phi_a(x)$ can be integrated out analytically, which leaves us with the partition function
\begin{equation}
  Z=\int\! DA_\mu D\psi D\bar{\psi}\, \exp\Bigg\{-\int\! d^4x\, \Big[\,\mathcal{L}_{\rm QCD} + \frac{4}{f_a^2}\, q(x)\,\Delta^{-1} q(x)\,\Big]\Bigg\}\,.
\end{equation}
For small values of $1/f_a^2$ the topological interaction term can be moved into the observable, leading to
\begin{equation}
  \int\! d^3\vec{x}\, \langle q(\vec{x},t)\,q(0) \rangle \simeq \int\! d^3\vec{x}\, \Big\langle q(\vec{x},t)\,q(0)\, \Big[1+\frac{4}{f_a^2}\int\! d^4y\; q(y)\,\Delta^{-1}q(y)\,\Big]\,\Big\rangle_{\rm QCD} \,,
\end{equation}
thus reducing the path integral to pure QCD. In QCD we expect $m_a = m_{\eta^\prime}$, at physical quark masses. And indeed,
\begin{equation}
- \lim_{t\rightarrow\infty} \frac{1}{t} \log \int\! d^3\vec{x}\, \langle q(\vec{x},t)\,q(0) \rangle_{\rm QCD}  = 1019\, (119)\, (^{+97}_{-86})\; \mbox{MeV} \,, 
\label{ma3}
\end{equation}
reported in~\cite{Fukaya:2015ara}, which is in good agreement with the experimental value $m_{\eta^\prime} = 958 \;\mbox{MeV}$. Approaching the weak coupling limit, we then get
\begin{equation}
  \begin{split}      
  m_a &= - \lim_{t\rightarrow\infty} \frac{1}{t} \log \int\! d^3\vec{x}\, \langle q(\vec{x},t)\,q(0) \rangle \\ &\rightarrow - \lim_{t\rightarrow\infty} \frac{1}{t} \log \int\! d^3\vec{x}\, \Big\langle q(\vec{x},t)\,q(0)\, \Big[1+\frac{4}{f_a^2}\int\! d^4y\; q(y)\,\Delta^{-1}q(y)\,\Big]\,\Big\rangle_{\rm QCD}\\ &\rightarrow - \lim_{t\rightarrow\infty} \frac{1}{t} \log \int\! d^3\vec{x}\, \langle q(\vec{x},t)\,q(0) \rangle_{\rm QCD} \, = \, m_{\eta^\prime} \,.
  \end{split}
\label{malim}  
\end{equation}
For cosmologically allowed couplings, $1/f_a^2 \lesssim 10^{-17}\, \mbox{GeV}^{-2}$, we should find $m_a = m_{\eta^\prime}$ within any conceivable numerical precision. 


The question is to what extent quantum effects will change the classical results. The answer to that demands a nonperturbative, `bottom-up'~\cite{Georgi:1994qn} evaluation of the theory. In this work we shall subject the Peccei-Quinn theory to a first quantitative test on the lattice.

\section{The QCD axion on the lattice}

At finite lattice spacing $a$ the `field theoretical' topological charge (\ref{topcharge}) is ill-defined~\cite{Luscher:1981zq} on the lattice. It happens that the $\theta$ term (\ref{thetaterm}) can be transformed into the quark mass matrix by an axial rotation, utilizing the axial anomaly~\cite{Baluni:1978rf,Guadagnoli:2002nm}. This has been exercised sucessfully in a recent lattice calculation of the electric dipole moment of the neutron~\cite{Guo:2015tla}. Similarly, the axion-gluon interaction in (\ref{axact2}) can be transformed into the fermionic part of the action by a space-time dependent axial rotation of the quark fields $\mathrm{q}$. Assuming three quark flavors, this is accomplished by
\begin{equation}
\mathrm{q}(x) \rightarrow \exp\left\{-i\gamma_5 \frac{c_{\mathrm{q}}}{2} \frac{\phi_a(x)}{f_a} \right\} \mathrm{q}(x)\,, \; \mathrm{q}=u,d,s\,,
\end{equation}
where $\displaystyle c_{\mathrm{q}}=\hat{m}/3m_{\mathrm{q}}$ and $\displaystyle \hat{m}^{-1}=\left(m_u^{-1} + m_d^{-1} + m_s^{-1}\right)/3$, $m_{\mathrm{q}}$ being the quark masses. Details of the transformation, including the most general axion action and its symmetries, is given in the Appendix. 

As a result of (\ref{fsize}), and for sufficiently large volumes $V$, we can expand the exponentials in the fermionic action (\ref{axactfff}) in powers of $\phi_a(x)/f_a$ and safely discard operators of dimension six and higher. This is even more the case for weak couplings, $1/f_a \rightarrow 0$, which is of particular interest in phenomenological applications. Replacing integrals by sums, we are then left with the lattice action 
\begin{equation}
S_{\rm Axion} =  a^4 \sum_x \,\Bigg[\,\frac{1}{2} \left(\partial_\mu \phi_a(x)\right)^2 - i\, \frac{\phi_a(x)}{3 f_a}\, \hat{m}\, \left(\bar{u}(x)\gamma_5 u(x) + \bar{d}\gamma_5 d(x) + \bar{s}(x)\gamma_5 s(x)\right)\Bigg]\,. 
\label{axactf}
\end{equation}  
The action (\ref{axactf}) is one of several totally equivalent QCD axion actions on the functional integral level. In the literature~\cite{Kim:1986ax} it is referred to as the Peccei-Quinn-Weinberg-Wilczek (PQWW) and Dine-Fischler-Srednicki-Zhitnitskii (DFSZ) parameterization of the action. Instantons are converted to zero modes by the axial anomaly, and fluctuations around it to low-lying, continuous modes of the Dirac operator. The equation of motion of this action is obtained from (\ref{eqmot}) by replacing $q(x)$ with $(\hat{m}/3)\,P(x)$, and leads directly to the expression (\ref{ma2}) for the axion mass. 

The action (\ref{axactf}) is complex and does not lend itself to numerical simulation on the Euclidean lattice. The finite volume partition function of the original action (\ref{CPact}) is analytic in $\theta$ for $|\theta|<\pi$~\cite{Aguado:2003ag}. And so are the resulting low-lying meson masses~\cite{Brower:2003yx,Acharya:2015pya}. For sufficiently large volumes, referring to (\ref{fsize}), we may expect that this holds for the action (\ref{axact2}) and parameter $1/f_a$ as well, so that we may resort to simulations at imaginary values of $f_a$, 
\begin{equation}
f_a^*=i f_a\,,
\end{equation}
being followed by analytic continuation to physical couplings. The final result is
\begin{equation}
S_{\rm Axion} = a^4 \sum_x \left[\frac{1}{2} \left(\partial_\mu \phi_a(x)\right)^2 + \frac{\hat{m}}{3} \frac{\phi_a(x)}{f_a^*}\, \left(\bar{u}(x)\gamma_5 u(x) + \bar{d}(x)\gamma_5 d(x) + \bar{s}(x)\gamma_5 s(x)\right)\right]\,.
\label{axactff}
\end{equation}   

We use the SLiNC action with Symanzik improved glue for QCD~\cite{Cundy:2009yy}. The action is nonperturbatively improved such that the axial Ward identity is satisfied. The kinetic part of the axion action is discretized as
\begin{equation}
\frac{1}{2}\, a^4 \sum_{x,\mu} \left(\frac{\phi_a(x+a\hat{\mu})-\phi_a(x)}{a}\right)^2 = \, a^2 \sum_{x,\mu} \left(\phi_a(x)-\phi_a(x+a\hat{\mu}\right)\phi_a(x)\,,
\end{equation}
$\hat{\mu}$ being the unit vector in $\mu$-direction. We assume periodic boundary conditions for axion and gauge fields. The gauge fields are updated using BQCD~\cite{Nakamura:2010qh}. Prior to the calculations reported in this paper we undertook a pilot study on the $24^3\times 48$ lattice, in which we simulated the full exponential $\exp\{c_{\rm q}\gamma_5\,\phi_a(x)/f_a^*\}$ in the fermionic action (\ref{axactfff}). We found no difference to the action (\ref{axactff}) for the parameters of Table~\ref{tab1}, as was to be expected.  

\begin{table}[t!]
\begin{center}
\begin{tabular}{c|c|c|c}
$\#$ & $a^{-4}V$ & $\kappa_0$ & $1/a f_a^*$ \\ \hline
1 & $12^3\times 24$ & 0.12090 & 0.01825 \\
2 & $12^3\times 24$ & 0.12090 & 0.1825\phantom{0} \\
3 & $12^3\times 24$ & 0.12090 & 1.825\phantom{00} \\ \hline
4 & $24^3\times 48$ & 0.12090 & 0.01825 \\
5 & $24^3\times 48$ & 0.12090 & 0.1825\phantom{0} \\
6 & $24^3\times 48$ & 0.12090 & 1.825\phantom{00} \\ \hline
7 & $32^3\times 64$ & 0.12090 & 0.1825\phantom{0} \\ 
\end{tabular}
\end{center}
\caption{The parameters of our present QCD + axion ensembles, generated at the SU(3) flavor symmetric point, for a wide range of bare axion decay constants, $2\kappa_0 \bar{m}/3a f_a^*=0.00001$, $0.0001$ and $0.001$, respectively. The parameters match previous pure QCD runs~\cite{Bietenholz:2011qq}.}
\label{tab1}
\end{table}

\begin{figure}[!t]
\vspace*{-1.25cm}
\begin{center}
\epsfig{file=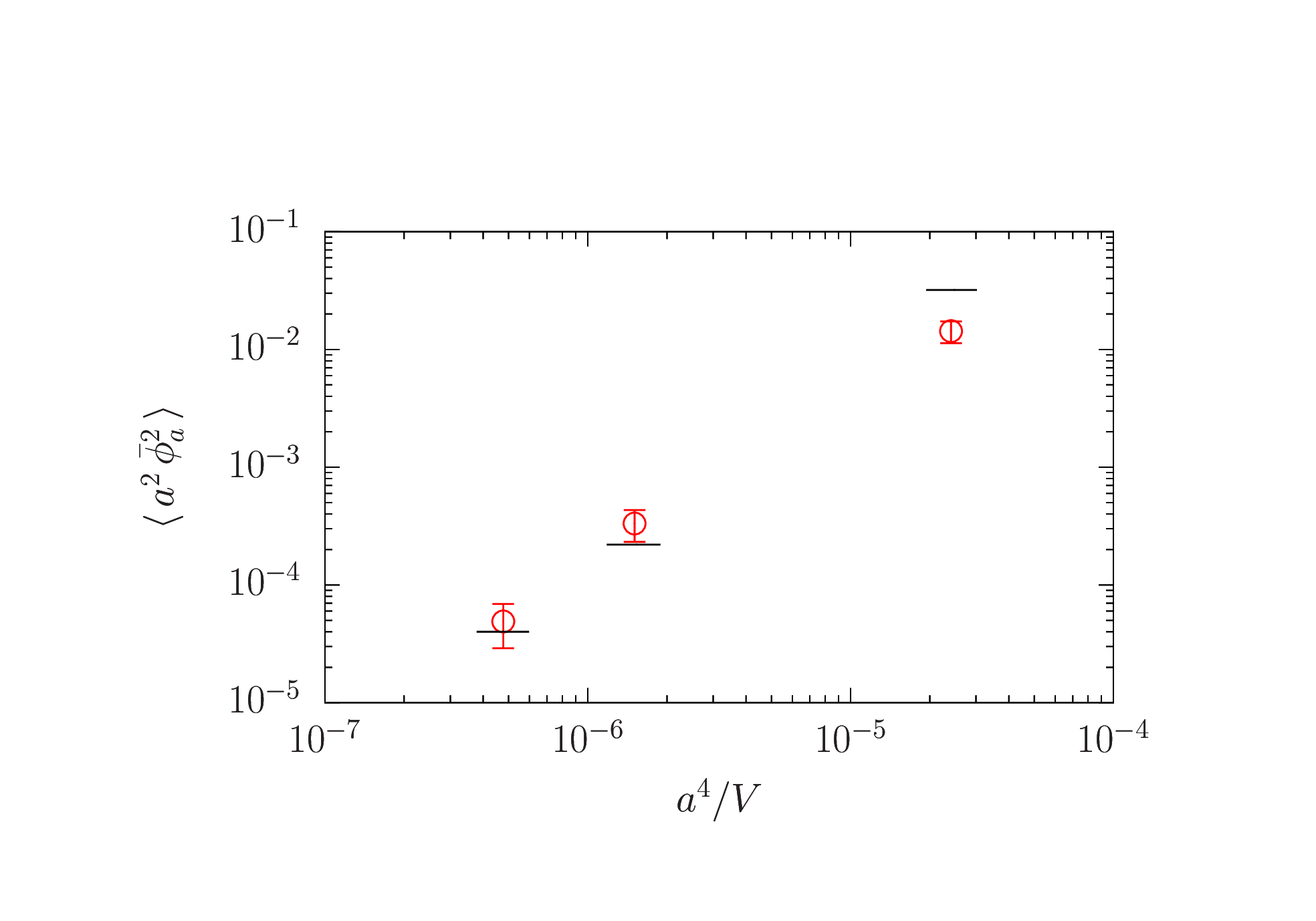,width=9.0cm,clip=} \hspace*{-2.0cm}
\epsfig{file=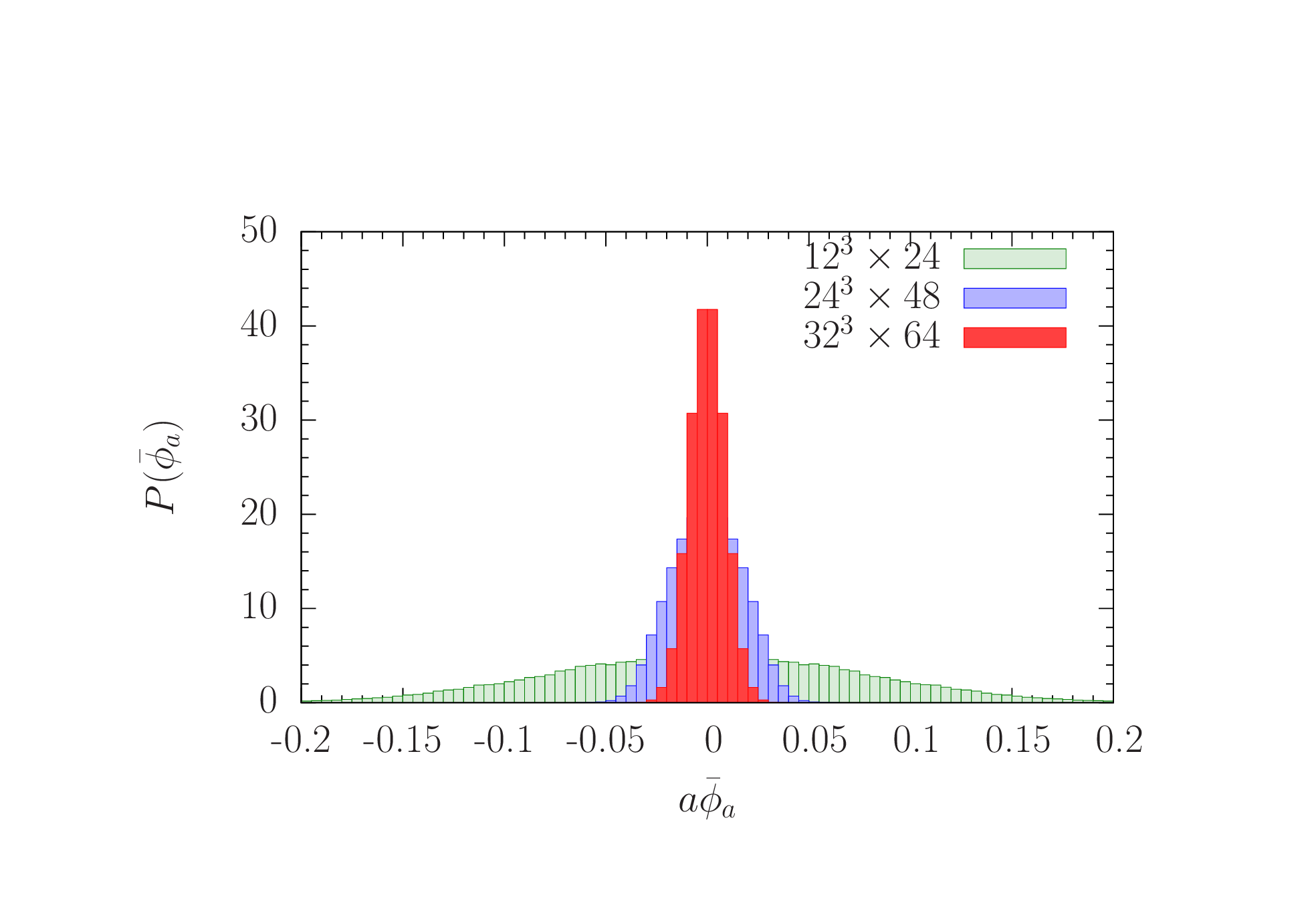,width=9.0cm,clip=}
\end{center}
\vspace*{-1.0cm}
\caption{Left: The average axion field squared $\langle\phibar_a^2\rangle$ as a function of volume for $1/a f_a^*=0.1825$. The horizontal bars represent $\langle \phibar_a^2\rangle = 1/m_a^2 V$, with $m_a$ taken from Table~\ref{tab2}. Errors are suppressed. Right: The probability distribution $P(\phibar_a)$ as a function of $\phibar_a$ and volume for $1/a f_a^*=0.1825$ as well.}
\label{fig0}
\end{figure}

As a first step, we focus on the SU(3) flavor symmetric point, defined by $m_u=m_d=m_s\equiv\bar{m}$ and $m_\pi^2=m_K^2=\left(m_\pi^{2\; {\rm phys}}+2m_K^{2\; {\rm phys}}\right)/3 = \left[415\,\mbox{MeV}\right]^2$, where we can hope for a strong signal, and which is a good starting point for simulations at smaller pion masses~\cite{Bietenholz:2011qq}. The simulations are done at $\beta=6/g^2 =5.50$ and restricted to $12^3\times 24$, $24^3\times 48$ and $32^3\times 64$ lattices, owing to the high computational cost. The quark mass $\bar{m}$ is given by $a\bar{m}=1/2\kappa_0-1/2\kappa_{0,c}$, where the hopping parameter $\kappa_0$ marks the symmetric point, and $\kappa_{0,c}$ is the critical hopping parameter at which $\bar{m}$ vanishes on the SU(3) symmetric line. To a good approximation $\kappa_0=0.12090$, while $\kappa_{0,c}=0.12110$. We use the center of mass of the nucleon octet to set the scale. This results in the lattice spacing $a=0.074(2)\,\mbox{fm}$~\cite{Bornyakov:2015eaa}. The simulation parameters are listed in Table~\ref{tab1}. Each ensemble consists of $O(10,000)$ configurations on the $12^3\times 24$ lattice, $O(1,500)$ configurations on the $24^3\times 48$ lattice and $O(1000)$ configurations on the $32^3\times 64$ lattice. 

We monitor $\phibar_a$ and the topological charge $Q$. In Fig.~\ref{fig0} we show $\langle\phibar_a^2\rangle$ and the probability distribution $P(\phibar_a)$ as a function of volume. As expected, $\langle \phibar_a^2 \rangle$ decreases rapidly with increasing volume. However, no significant dependence of $\langle \phibar_a^2 \rangle$ on $1/f_a$ is observed. On the $24^3\times 48$ lattice, for example, $\langle a^2 \phibar_a^2 \rangle = 1.3 \cdot 10^{-4}$ and $1.8 \cdot 10^{-4}$ for $1/af_a^* = 0.1825$ and $0.01825$, respectively. Instead, the data appear to follow (\ref{fsize}) but with $(\chi_t/f_a^2 + \mu_a^2)$ replaced by the actual mass squared, $\langle \phibar_a^2 \rangle = 1/m_a^2 V$. Furthermore, $P(\phibar_a)$ turns out to be well described by a Gaussian distribution, which shrinks to zero equally rapidly. On the $24^3\times 48$ and $32^3\times 64$ lattices $\langle \phibar_a^2\rangle/f_a^2$ ranges from $7.3 \cdot 10^{-6}$ to $1.3 \cdot 10^{-6}$, assuring that contributions of higher powers of $\phi_a/f_a$ to the action can indeed be neglected.

\begin{figure}[!b]
\vspace*{-1.0cm}
\begin{center}
\epsfig{file=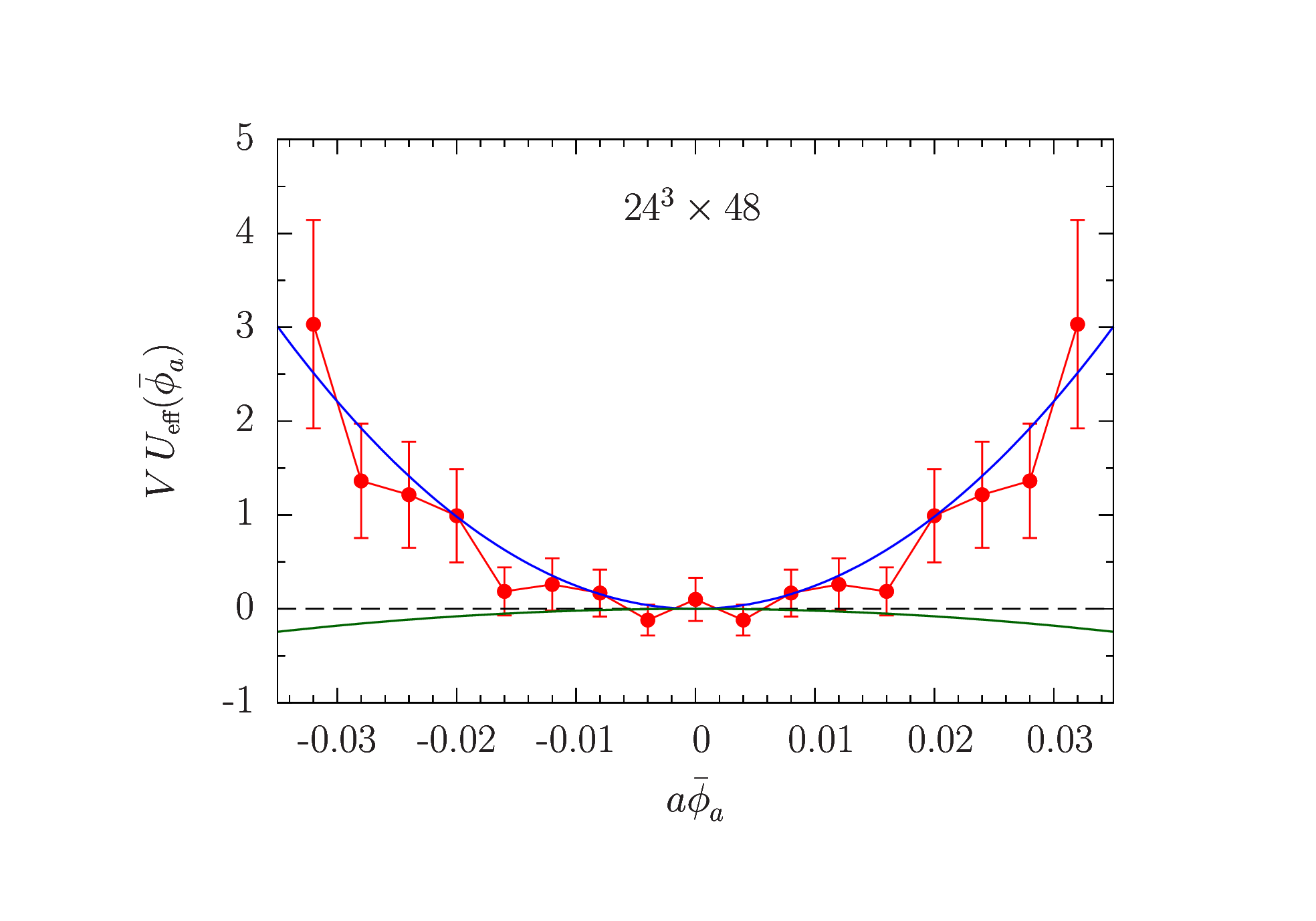,width=12cm,clip=}
\end{center}
\vspace*{-1.0cm}
\caption{The effective potential $U_{\rm eff}(\phibar_a)$ on the $24^3\times 48$ lattice for $1/a f_a^*=0.01825$. The potential is normalized to zero at $\phibar_a=0$. The upper (blue) curve shows a fit to the lattice data. The solid bottom (green) curve represents the classical potential (\ref{classpot}) for imaginary $f_a = i f_a^*$.}
\label{figeffpot}
\end{figure}

In order that the Hamiltonian is bounded from below, the effective potential must be positive. For imaginary values of $f_a$ this is not the case for the classical potential (\ref{classpot}), which changes sign under the transformation $1/f_a^2 \rightarrow - 1/f_a^2$. The effective potential (\ref{effpotphi}), in contrast, turns out to be positive and well behaved. In Fig.~\ref{figeffpot} we show $U_{\rm eff}(\phibar_a)$ for the $24^3\times 48$ lattice at $1/a f_a^*=0.01825$, together with a quadratic fit of the form $U_{\rm eff}(\phibar_a) = (m_a^2/2)\,\phibar_a^2$. The result of the fit is $m_a = 230(60)\,\mbox{MeV}$, which is consistent with the lattice result given in Table~\ref{tab2}. 
Similar results are found for the other $24^3\times 48$ and $32^3\times 64$ lattices. 

\begin{figure}[!b]
\vspace*{-1.25cm}
\begin{center}
\epsfig{file=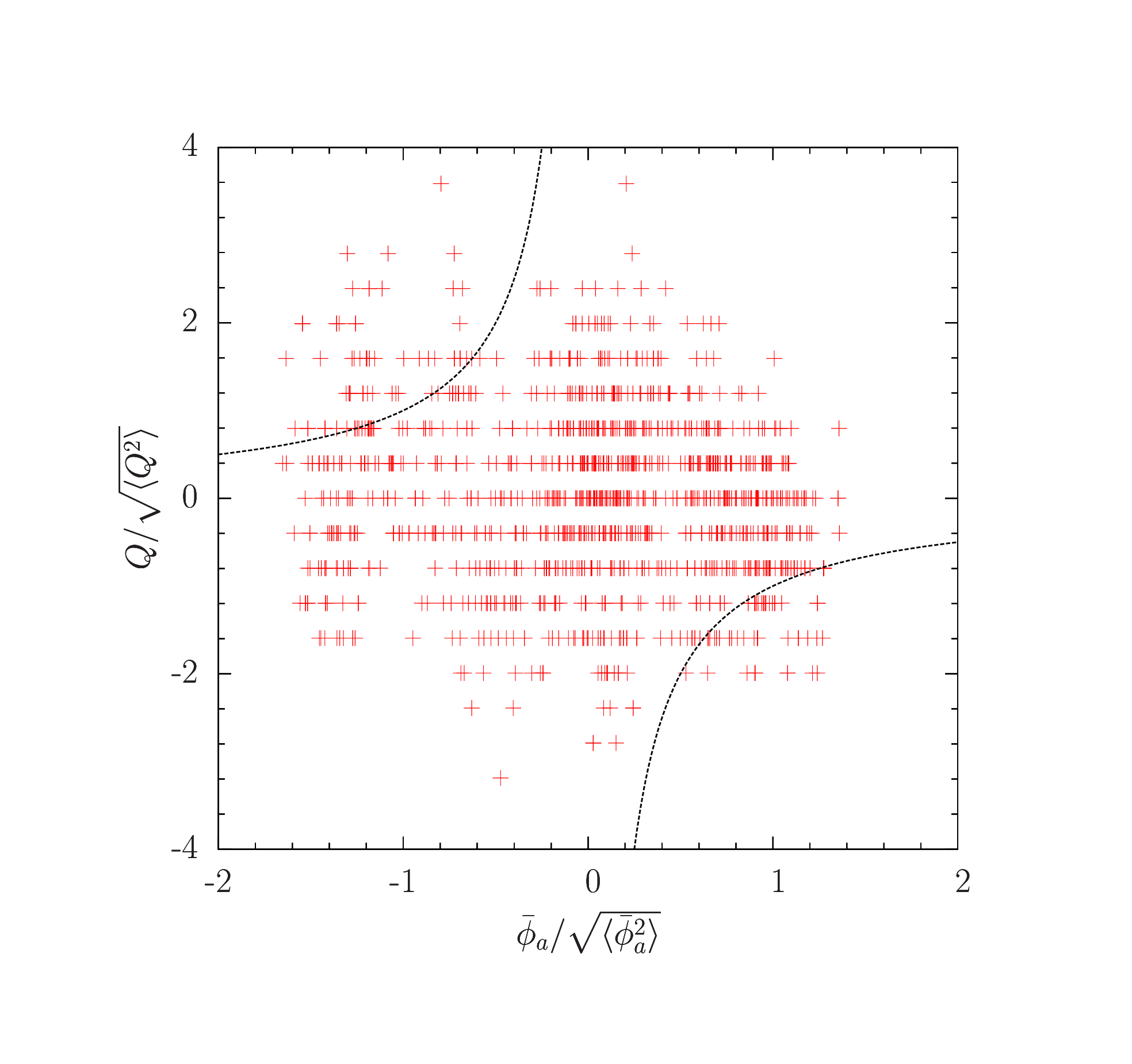,width=12cm,clip=} 
\end{center}
\vspace*{-1.0cm}
\caption{The topological charge $Q$ obtained from cooling~\cite{cooling} plotted against the average axion field $\phibar_a$, configuration by configuration, on the $24^3\times 48$ lattice at $1/a f_a^*=0.01825$. The solid curves correspond to the classical result $r_{\phibar_a Q}=-1$.}
\label{fig1}
\end{figure}

The classical theory predicts that $\phibar_a$ and the topological charge $Q$ are strongly correlated. A measure for linear relationship is Pearson's correlation coefficient
\begin{equation}
r_{\phibar_a Q}=\frac{\sum_1^N\,\phibar_a Q - \sum_1^N \phibar_a \sum_1^N Q/N}{\sqrt{\left(\sum_1^N \phibar_a^2-\left(\sum_1^N \phibar_a\right)^2\!\!/N\right)\left(\sum_1^N Q^2-\left(\sum_1^N Q\right)^2\!\!/N\right)}} \,,
\end{equation}
where $\sum_1^N$ is the ensemble sum. From the partion function $Z$ with $\mu_a=0$ we derive 
\begin{equation}
r_{\phibar_a Q} = -1\,,
\label{pearson}
\end{equation}
which means that $Q$ and $\phibar_a$ would be exactly anticorrelated, to be precise. In Fig.~\ref{fig1} we show a scatter plot of $Q$ versus $\phibar_a$, together with the classical prediction (\ref{pearson}). If (\ref{pearson}) were correct, the lattice data would cluster around the solid curves. Instead, the lattice data do not show any correlation between $Q$ and $\phibar_a$. On the $24^3\times 48$ lattice shown in the figure we find $r_{\phibar_a Q}=0.016(113)$, which is compatible with zero. Similar results are found for the other lattices. This confirms that the axion field takes little to no account of the global topological charge $Q$, as we have reasoned before. With $\mu_a=m_a$, we derive $r_{\phibar_a Q} = -1/(1+m_a^2 f_a^2/\chi_t)$. This is consistent with $r_{\phibar_a Q} = 0$ to several decimal places for the parameters in Table~\ref{tab2}, and fits our lattice results.

\section{The axion mass}

\begin{figure}[!b]
\vspace*{-0.65cm}
\begin{center}
\epsfig{file=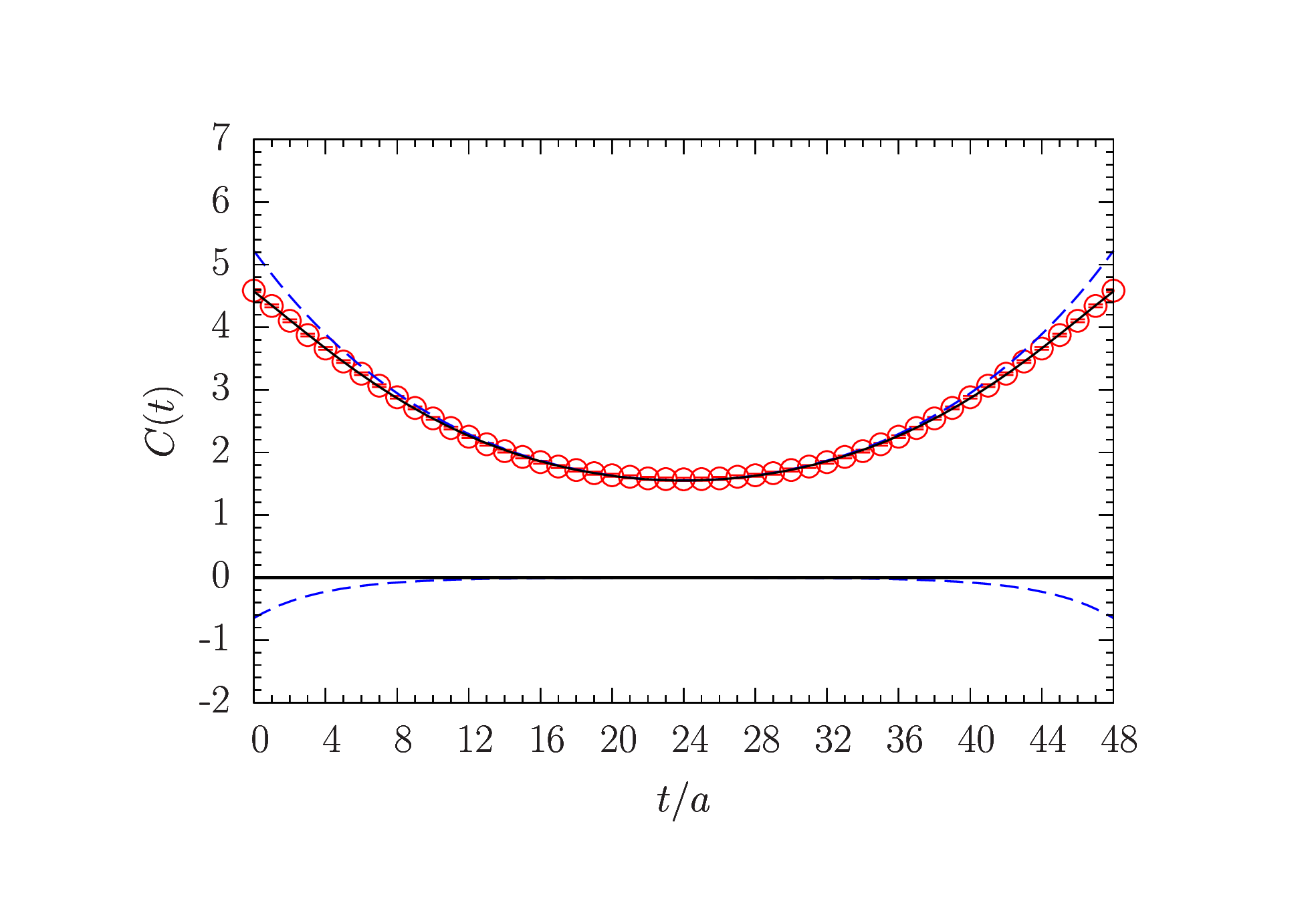,width=9cm,clip=} \hspace*{-2.0cm}
\epsfig{file=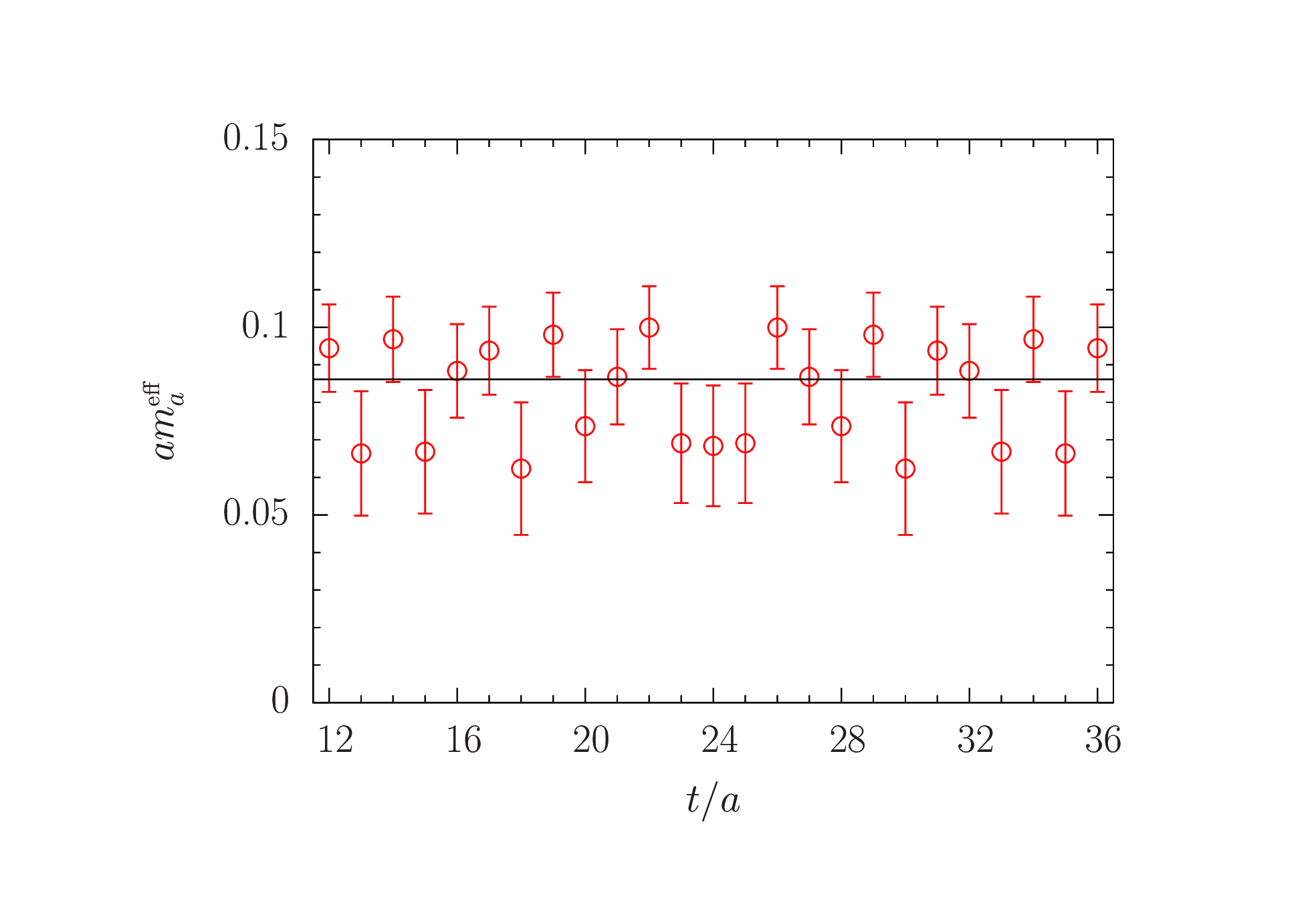,width=9cm,clip=}
\end{center}
\vspace*{-1.0cm}
\caption{The correlation function $C(t)$ (left) and the effective axion mass $m_a^{\rm eff}$ (right) on the $24^3\times 48$ lattice for $1/a f_a^*=0.01825$. The upper dashed curve (left) shows the contribution of the axion, $A \cosh\big(-m_a \tau\big)$, while the lower dashed curve shows the contribution of the $\eta^\prime$, $B \cosh\big(-m_{\eta^\prime} \tau\big)$.}
\label{fig3}
\end{figure}

\begin{figure}[!b]
\vspace*{-1.75cm}
\begin{center}
\epsfig{file=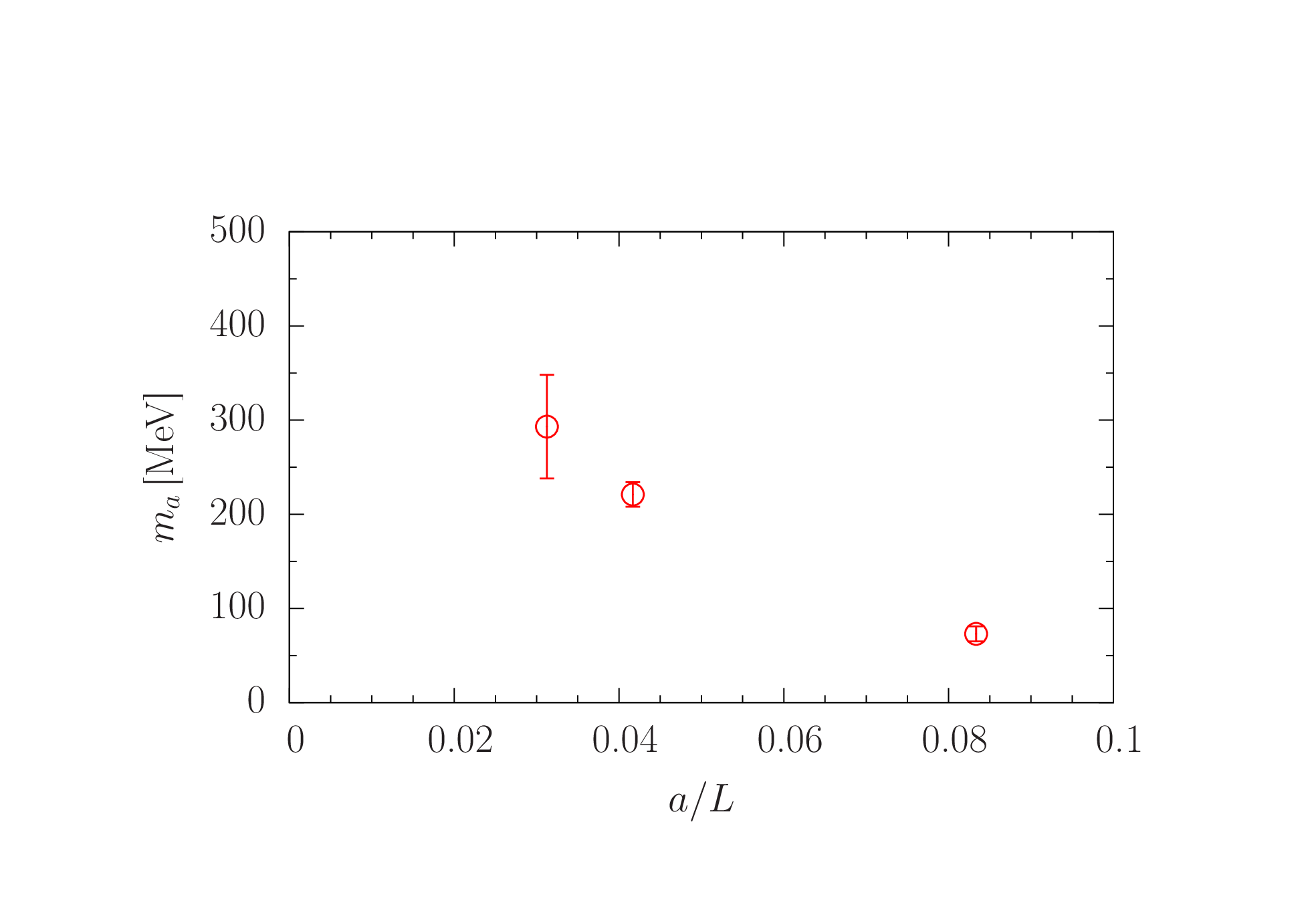,width=12cm,clip=}
\end{center}
\vspace*{-1.0cm}
\caption{The axion mass $m_a$ for $1/ f_a^* = 0.068\, \mbox{GeV}^{-1}$ as a function of the spatial extent of the lattice.}
\label{fig4}
\end{figure}

To determine the axion mass $m_a$, we compute the correlation function
\begin{equation}
C(t)=a^2 \sum_{\vec{x}}\, \big(\langle \phi_a(\vec{x},t) \, \phi_a(0) \rangle - \langle \phibar_a^2 \rangle \big)\,,\\[-0.75em]
\label{aa}
\end{equation}
which we parameterize as
\begin{equation}
C(t)= A \cosh\big(-m_a \tau\big) + B \cosh\big(-m_{\eta^\prime} \tau\big) \,, \; \tau = t-T/2 \,,
\label{corrfu}
\end{equation}
where $T$ is the temporal extent of the lattice. Note that any exactly massless particle is not detected by the correlation function (\ref{aa}). In addition to the axion, which we assume to be the lightest particle, we expect the operator $\phi_a$ to couple to a flavor singlet quark-antiquark bound state, which we call the $\eta^\prime$ meson. In Fig.~\ref{fig3} we show the correlation function $C(t)$ on the $24^3\times 48$ lattice together with a two-exponential fit of (\ref{corrfu}) to the data. The individual contributions of the axion and the $\eta^\prime$ are shown by the dashed curves. As expected, the contribution of the $\eta^\prime$ is negative, as a result of reflection positivity. Also shown is the effective mass of the axion,
\begin{equation}
am_a^{\rm eff} = {\rm arcosh}\big[\big(C(t-a)+C(t+a)\big)/2C(t)\big]\,,
\end{equation}
which dominates the correlation function at times $8 << t/a << 40$. In Table~\ref{tab2} we collect our results for the axion and $\eta^\prime$ masses. The $12^3\times 24$ lattice was too small and the statistics of the $32^3\times 64$ lattice too low to determine the $\eta^\prime$ mass reliably. The axion mass shows significant finite size corrections. In Fig.~\ref{fig4} we plot the axion mass against the spatial extent of the lattice for $1/ f_a^* = 0.068\, \mbox{GeV}^{-1}$. In the infinite volume limit $m_a$ tends to approach $\approx 400 \,\mbox{MeV}$, which appears to coincide with the mass of the $\eta$ meson and the pion at the SU(3) flavor symmetric point~\cite{Bietenholz:2011qq}, $m_\eta = m_\pi = 415\, \mbox{MeV}$. The $\eta^\prime$ masses of Table~\ref{tab2} are consistent with the expected result at the SU(3) flavor symmetric point~\cite{Bali:2017qce}. The topological susceptibility $\chi_t$ turns out to be in agreement with pure QCD at the given mass, but shows no dependence on $1/f_a^*$. Finally, it should be noted that the axion masses in Table~\ref{tab2}, computed from the correlation function (\ref{corrfu}), are in total agreement with the masses from a quadratic fit, $U_{\rm eff}(\phibar_a)=(m_a^2/2)\, \phibar_a^2$, to the effective potential $V\,U_{\rm eff}(\phibar_a)= - \log P(\phibar_a) + c$, as shown in Fig.~\ref{figeffpot}.

\begin{table}[!t]
\vspace*{0.25cm}
\begin{center}
\begin{tabular}{c|c|c|c|c|c}
$\#$ & $a^{-4}V$ & $1/ f_a^* \, [\mbox{GeV}^{-1}]$ & $\chi_t^{1/4} \, \mbox{[MeV]}$ &  $m_a \, \mbox{[MeV]}$ & $m_{\eta^\prime} \, \mbox{[MeV]}$ \\ \hline
1 & $12^3\times 24$ & 0.0068 & $119 \pm \!\phantom{0}4$ & $\phantom{0}62 \pm \!\phantom{0}2$ &\\
2 & $12^3\times 24$ & 0.068\phantom{0} & $121 \pm \!\phantom{0}6$ & $\phantom{0}73 \pm \!\phantom{0}8$ &\\
3 & $12^3\times 24$ & 0.68\phantom{00} & $108 \pm \!\phantom{0}8$ & $\phantom{0}66 \pm \!\phantom{0}4$ &\\ \hline
4 & $24^3\times 48$ & 0.0068 & $153 \pm \!11$ & $230 \pm \!13$ & $700 \pm 110$\\[0.1em]
5 & $24^3\times 48$ & 0.068\phantom{0} & $148 \pm \!\phantom{0}9$ & $221 \pm \!13$ & $660^{\displaystyle\: + \,\phantom{0}50}_{\displaystyle\: -\, 350}$\\[0.5em]
6 & $24^3\times 48$ & 0.68\phantom{00} & $151 \pm \!\phantom{0}8$ & $238 \pm \!11$ & $670 \pm 120$\\ \hline
7 & $32^3\times 64$ & 0.068\phantom{0} & $152 \pm \!\phantom{0}9$ & $293 \pm \!55$ & \\
\end{tabular}
\end{center}
\caption{The axion and $\eta^\prime$ masses, as well as the topological susceptibility, according to volume and bare axion decay constant $f_a^*$, in physical units using $a=0.074(2)\, \mbox{fm}$.}
\label{tab2}
\end{table}

We do not find any signal of a light axion for values of $f_a^*$ ranging from approximately $1.5$ to $150 \,\mbox{GeV}$. Making use of the equation of motion and the axial anomaly, the correlator (\ref{aa}) can be re-expressed as
\begin{equation}
C(t) = \langle P(t)\, P(0)\rangle_{disc}  
\end{equation}
up to a multiplicative constant, where $P=(\bar{u}\gamma_5 u +\bar{d}\gamma_5 d +\bar{s}\gamma_5 s)$. At the SU(3) flavor symmetric point, which we exclusively deal with in this paper, C(t) decomposes into flavor octet and flavor singlet correlation functions $C_{88}$ and $C_{11}$~\cite{Bali:2017qce},
\begin{equation}
C(t) = \frac{1}{3} \left( C_{88}(t) - C_{11}(t)\right) \,. 
\end{equation}
In pure QCD $C_{88}(t) = C_\eta(t) \propto \exp\{-m_\eta\, t\}$ and $C_{11}(t) = C_{\eta^\prime}(t) \propto \exp\{-m_{\eta^\prime}\,t\}$, which suggests that $m_a \rightarrow m_\eta \approx 400 \,\mbox{MeV}$ as $1/f_a \rightarrow 0$. 
This is exactly what we find, a lighter particle approaching the mass of the $\eta$ and a heavier particle with negative correlation function being identified as the $\eta^\prime$ meson. A very light axion, as predicted by the classical theory, would have shown up in a very shallow effective potential and in much larger values of $\langle \phibar_a^2\rangle$. 

\begin{figure}[!t]
\vspace*{-2.5cm}
\begin{center}
\epsfig{file=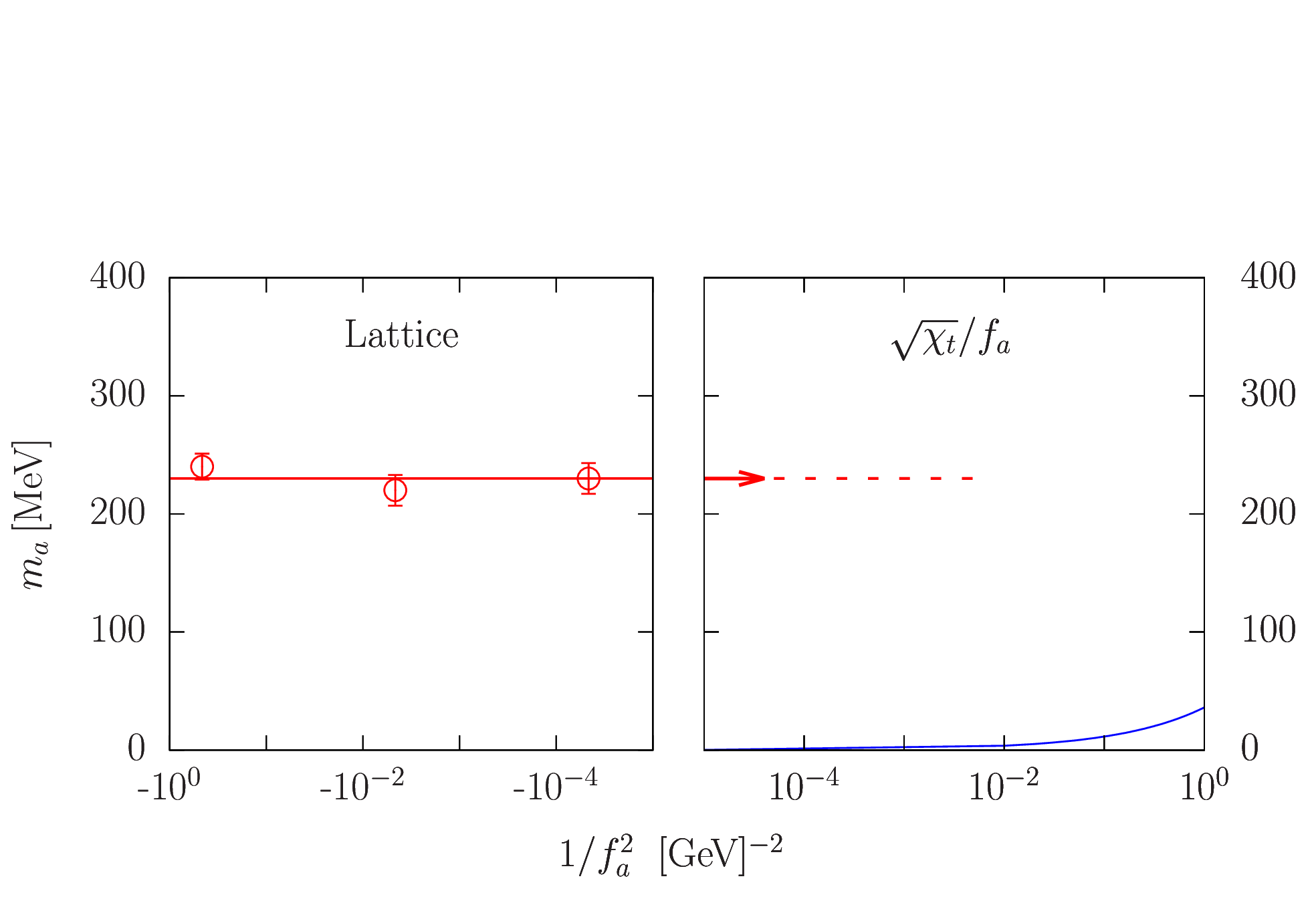,width=14cm,clip=}
\end{center}
\vspace*{-0.75cm}
\caption{The axion mass $m_a$ as a function of $1/f_a^2$ on the $24^3\times 48$ lattice (left), compared with the prediction (\ref{axmass}) of the semi-classical approximation (right).}
\label{fig5}
\end{figure}

From the very beginning we argued that the axion mass $m_a$, whatever the interpretation is, should turn out to be more or less independent of $f_a$ for sufficiently small values of $1/f_a$. Our lattice calculations confirm that. In Fig.~\ref{fig5} we show the axion mass from the $24^3\times 48$ lattice as a function of $1/f_a^2$. The equation of motion (\ref{eqmot}) and mass formulae (\ref{ma1}), (\ref{ma2}) and (\ref{malim}) suggest that the axion mass continues more ore less linearly across $1/f_a^2 = 0$ to positive values of $1/f_a^2$, as indicated in Fig.~\ref{fig5}. The point $1/f_a^2=0$ is an exceptional point, with $m_a$ being analytic all around it.

\section{Conclusion and outlook}

The strong CP problem is a nonperturbative problem, which requires nonperturbative techniques to solve it. In this work we have performed, for the first time, a fully dynamical simulation of the QCD axion on the lattice for three flavors of quarks and axion decay constants ranging from approximately $1.5$ to $150\,\mbox{GeV}$. 
No light axion has been found, which would qualify as a candidate for dark matter. Instead we found that the axion combines and mixes with the pseudoscalar meson sector. The result was a lighter particle, with mass $m_a$ approaching the mass of the $\eta$ meson in the infinite volume, and a heavier particle, which we identify as the $\eta^\prime$ meson. This tallies with our expectations based on general principles, such as the equation of motion and axial anomaly. The results of the calculation have been contrasted with the predictions of the effective theory, for which we do not find much support. 
The physical situation here is similar to that of the $\eta^\prime$ meson, which is not a Goldstone boson either, nor can its mass be accounted for by effective theory.

The clue is that the lattice topological charge density $q(x)$ defies any semi-classical interpretation. Instantons provide only zeroth-order contributions to path integrals. While the distribution of total charge $Q$ is well described by a dilute gas of instantons, $q(x)$ turns out to be dominated by far by low-lying, continuous eigenmodes of the Dirac operator~\cite{Ilgenfritz:2005hh}, which arise from quantum fluctuations. To account for these fluctuations analytically, one would have to compute the functional integral about the multi-instanton background field, the instanton determinant, which has resisted any solution so far. 

Our calculations have been done at the SU(3) flavor symmetric point. At this point $\eta$ and pion are pure octet states and equal in mass, while the $\eta^\prime$ is a pure singlet state. Away from the symmetric point $\eta$ and $\eta^\prime$ will mix. To reveal the full mixing matrix in the most general case one needs to solve for the generalized eigenvalues of the correlation matrix involving the axion and $\eta$ and $\eta^\prime$ states. In the limit of physical quark masses and weak coupling we expect the axion mass to approach the mass of the $\eta^\prime$.  

The simulations became feasible by rotating the anomalous axion-gluon coupling into the fermion mass matrix. The exponential $\exp\{-i c_{\rm q} \gamma_5 \phi_a/f_a\}$ in (\ref{axactfff}) could be expanded about $\phi_a/f_a = 0$ due to the rather small values of $\phi_a/f_a$, limiting the action to operators of dimension five and lower. For nonvanishing vacuum angle $\theta$ this is no longer true. In this case we are faced with the exponential $\exp\{-i c_{\rm q} \gamma_5 (\phi_a/f_a+\theta)\}$. Furthermore, $(\phibar_a/f_a)\,Q$ in (\ref{part}) needs to be replaced by $(\phibar_a/f_a +\theta)\,Q$, which leads to  
\begin{equation}
\langle \phibar_a^2 \rangle = \frac{1}{(\chi_t/f_a^2 + \mu_a^2)\, V} + f_a^2\, \theta^2 \,,
\end{equation}
indicating that $\langle \phibar_a^2 \rangle$ will cluster around $f_a^2\,\theta^2$. This suggest an expansion about $\phi_a = - f_a\, \theta$. Writing $\phi_a(x) = -f_a\, \theta + \phi_a^\delta(x)$ and replacing $\phi_a(x)$ by $\phi_a^\delta(x)$ in our simulations should give the same results for the masses, maintaining the shift symmetry and thus removing the CP violating $\theta$ term. This needs to be checked though. 
 
So far the calculations have been done at a single value of the lattice spacing, $a = 0.074\,\mbox{fm}$. It is an open question, what happens at larger momentum cut-offs, and whether the QCD axion can be a self-consistent, fundamental quantum field theory with a well defined continuum limit. Simulations at varying lattice spacings, together with analytical investigations~\cite{Eichhorn:2012uv}, will have to show. 

\section{Appendix}

The most general Lagrangian for the axion field $\phi_a$ up to order $1/f_a$ is~\cite{Weinberg}
\begin{equation}
\mathcal{L}_{\phi_a} = \frac{1}{2} (\partial_\mu \phi_a(x))^2 - i \frac{\phi_a(x)}{f_a}\, \frac{1}{64\pi^2}\,\epsilon_{\mu\nu\rho\sigma} F_{\mu\nu}^a(x) F_{\rho\sigma}^a(x) + i f \frac{\partial_\mu \phi_a(x)}{f_a}\, A_\mu\,,
\end{equation}
where $A_\mu$ is the (flavor singlet) axial vector current and $f$ a dimensionless coupling constant. By a redefinition of the quark fields
\begin{equation}
\mathrm{q}(x) \rightarrow \exp\left\{-i \alpha \gamma_5 \frac{c_{\mathrm{q}}}{2} \frac{\phi_a(x)}{f_a}\right\} \mathrm{q}(x) \,,\; \mathrm{q} =u,d,s
\label{pqtrans}
\end{equation}
the measure of the quark fields changes by
\begin{equation}
[d\mathrm{q}][d\bar{\mathrm{q}}] \rightarrow \exp\left\{- i \alpha c_{\mathrm{q}} \frac{\phi_a}{f_a}\, \frac{1}{64\pi^2}\, \epsilon_{\mu\nu\rho\sigma} F_{\mu\nu}^a F_{\rho\sigma}^a\right\}[d\mathrm{q}][d\bar{\mathrm{q}}]\,.
\end{equation}
As a result, the topological interaction term
\begin{equation}
- i \frac{\phi_a(x)}{f_a} \, \frac{1}{64\pi^2}\,\epsilon_{\mu\nu\rho\sigma} F_{\mu\nu}^a(x) F_{\rho\sigma}^a(x)
\label{topterm}
\end{equation}
can be moved into the fermionic part of the Lagrangian. In its most general form, the Lagrangian for $\phi_a$ can then be written
\begin{equation}
\begin{split}
\mathcal{L}_{\phi_a} &=\frac{1}{2} (\partial_\mu \phi_a(x))^2 + m_u\, \bar{u}(x)\, \exp\left\{-i c_1 \gamma_5\, c_u \frac{\phi_a(x)}{f_a}\right\} u(x) +
m_d\, \bar{d}(x)\, \exp\left\{-i c_1 \gamma_5\, c_d \frac{\phi_a(x)}{f_a}\right\} d(x) \\ &+
m_s\, \bar{s}(x)\, \exp\left\{-i c_1 \gamma_5\, c_s \frac{\phi_a(x)}{f_a}\right\} s(x) + i c_2 \frac{\phi_a(x)}{f_a} \, \frac{1}{64\pi^2}\,\epsilon_{\mu\nu\rho\sigma} F_{\mu\nu}^a(x) F_{\rho\sigma}^a(x) + i c_3 \frac{\partial_\mu \phi_a(x)}{f_a}\, A_\mu\,,
\end{split}
\label{axactfff}
\end{equation}
where $c_1=\alpha$, $c_2=1-\alpha$ and $c_3=f-\alpha/2$. The contribution to $c_3$ arises from the kinematic part of the QCD Lagrangian. For $\alpha=1$ the topological interaction term (\ref{topterm}) is totally eliminated. 

The KSVZ, PQWW and DFSZ actions mentioned in the main text refer to
\begin{equation}
\begin{tabular}{llll}
KSVZ: & $c_1 = 0\,,$ & $c_2 \neq 0\,,$ & $c_3 = 0\,,$\\
PQWW: & $c_1\neq 0\,,$ & $c_2 = 0\,,$ & $c_3 = 0\,,$\\
DFSZ: & $c_1\neq 0\,,$ & $c_2 = 0\,,$ & $c_3 = 0\,.$
\end{tabular}
\end{equation}
In all three cases the derivative interaction has been discarded ($c_3=0$). The axion mass depends only on the combination $c_1+c_2$~\cite{Kim:1986ax}. Under an arbitrary transformation (\ref{pqtrans}) the coefficients $c_1$, $c_2$ and $c_3$ transform as $c_1 \rightarrow c_1 - \alpha$, $c_2 \rightarrow c_2 + \alpha$ and $c_3 \rightarrow c_3 - \alpha/2$,
which leaves the combination $c_1+c_2$, and thus the axion mass, invariant.  

In this work we shall use the PQWW/DFSZ parameterization of the action, with $c_1=1$ and $c_2=c_3=0$. The action (\ref{axactf}) is obtained by expanding (\ref{axactfff}) to leading order in $1/f_a$, that is discarding operators of dimension six and higher.


\begin{thebibliography}{99}

\bibitem{Baker}
  C.A.~Baker {\it et al.},
  Phys.\ Rev.\ Lett.\  {\bf 97} (2006) 131801
  [arXiv:hep-ex/0602020];
  C.A.~Baker {\it et al.},
  Phys.\ Rev.\ Lett.\  {\bf 98} (2007) 149102 
  [arXiv:0704.1354 [hep-ex]].

\bibitem{Guo:2015tla}
  F.-K.~Guo, R.~Horsley, U.-G.~Mei{\ss}ner, Y.~Nakamura, H.~Perlt, P.E.L.~Rakow, G.~Schierholz, A.~Schiller and J.M.~Zanotti,
  Phys.\ Rev.\ Lett.\  {\bf 115} (2015) 062001
  [arXiv:1502.02295 [hep-lat]].

\bibitem{Peccei:1977hh}
  R.D.~Peccei and H.R.~Quinn,
  Phys.\ Rev.\ Lett.\  {\bf 38} (1977) 1440,
  Phys.\ Rev.\ D {\bf 16} (1977) 1791.

\bibitem{Kim:1986ax}
  J.E.~Kim,
  Phys.\ Rept.\  {\bf 150} (1987) 1;
  J.E.~Kim and G.~Carosi,
  Rev.\ Mod.\ Phys.\  {\bf 82} (2010) 557
  [arXiv:0807.3125 [hep-ph]].

\bibitem{DelDebbio:2004ns}
  L.~Del Debbio, L.~Giusti and C.~Pica,
  Phys.\ Rev.\ Lett.\  {\bf 94} (2005) 032003
  [hep-th/0407052].

\bibitem{Peccei:2006as}
  R.D.~Peccei,
  Lect.\ Notes Phys.\  {\bf 741} (2008) 3
  [hep-ph/0607268].

\bibitem{Weinberg:1977ma}
  S.~Weinberg,
  Phys.\ Rev.\ Lett.\  {\bf 40} (1978) 223.

\bibitem{Wilczek:1977pj}
  F.~Wilczek,
  Phys.\ Rev.\ Lett.\  {\bf 40} (1978) 279.

\bibitem{Fukuda:1974ey}
  R.~Fukuda and E.~Kyriakopoulos,
  Nucl.\ Phys.\ B {\bf 85} (1975) 354.

\bibitem{ORaifeartaigh:1986axd}
  L.~O'Raifeartaigh, A.~Wipf and H.~Yoneyama,
  Nucl.\ Phys.\ B {\bf 271} (1986) 653.

\bibitem{Ilgenfritz:2005hh}
  E.-M.~Ilgenfritz, K.~Koller, Y.~Koma, G.~Schierholz, T.~Streuer and V.~Weinberg,
  Nucl.\ Phys.\ Proc.\ Suppl.\  {\bf 153} (2006) 328
  [hep-lat/0512005].

\bibitem{Ilgenfritz:2007xu}
  E.-M.~Ilgenfritz, K.~Koller, Y.~Koma, G.~Schierholz, T.~Streuer and V.~Weinberg,
  Phys.\ Rev.\ D {\bf 76} (2007) 034506
  [arXiv:0705.0018 [hep-lat]].

\bibitem{Hasenfratz:1998ri}
  P.~Hasenfratz, V.~Laliena and F.~Niedermayer,
  Phys.\ Lett.\ B {\bf 427} (1998) 125
  [hep-lat/9801021].

\bibitem{Fukaya:2015ara}
  H.~Fukaya, S.~Aoki, G.~Cossu, S.~Hashimoto, T.~Kaneko and J.~Noaki,
  Phys.\ Rev.\ D {\bf 92} (2015) 111501
  [arXiv:1509.00944 [hep-lat]].  
  
\bibitem{Georgi:1994qn}
  H.~Georgi,
  Ann.\ Rev.\ Nucl.\ Part.\ Sci.\  {\bf 43} (1993) 209.

\bibitem{Luscher:1981zq}
  M.~L\"uscher,
  Commun.\ Math.\ Phys.\  {\bf 85} (1982) 39.

\bibitem{Baluni:1978rf}
  V.~Baluni,
  Phys.\ Rev.\ D {\bf 19} (1979) 2227.

\bibitem{Guadagnoli:2002nm}
  D.~Guadagnoli, V.~Lubicz, G.~Martinelli and S.~Simula,
  JHEP {\bf 0304} (2003) 019
  [hep-lat/0210044].

\bibitem{Aguado:2003ag}
  M.~Aguado, M.~Asorey and D.~Garcia-Alvarez,
  Mod.\ Phys.\ Lett.\ A {\bf 18} (2003) 2303;
  M.~Aguado and M.~Asorey,
  AIP Conf.\ Proc.\  {\bf 1343} (2011) 173.

\bibitem{Brower:2003yx}
  R.~Brower, S.~Chandrasekharan, J.W.~Negele and U.-J.~Wiese,
  Phys.\ Lett.\ B {\bf 560} (2003) 64
  [hep-lat/0302005].

\bibitem{Acharya:2015pya}
  N.R.~Acharya, F.-K.~Guo, M.~Mai and U.-G.~Mei{\ss}ner,
  Phys.\ Rev.\ D {\bf 92} (2015) 054023
  [arXiv:1507.08570 [hep-ph]].


\bibitem{Cundy:2009yy}
  N.~Cundy. M.~G\"ockeler, R.~Horsley, T.~Kaltenbrunner, A.D.~Kennedy, Y.~Nakamura, H.~Perlt, D.~Pleiter, P.E.L.~Rakow, A.~Sch\"afer, G.~Schierholz, A.~Schiller, H.~St\"uben and J.M.~ Zanotti,
  Phys.\ Rev.\ D {\bf 79} (2009) 094507
  [arXiv:0901.3302 [hep-lat]].

\bibitem{Nakamura:2010qh}
  Y.~Nakamura and H.~St\"uben,
  PoS LATTICE {\bf 2010} (2010) 040
  [arXiv:1011.0199 [hep-lat]].

\bibitem{Bietenholz:2011qq}
  W.~Bietenholz, V.~Bornyakov, M.~G\"ockeler, R.~Horsley, W.G.~Lockhart, Y.~Nakamura, H.~Perlt, D.~Pleiter, P.E.L.~Rakow, G.~Schierholz, A.~Schiller, T.~Streuer, H.~St\"uben, F.~Winter and J.M.~Zanotti,
  Phys.\ Rev.\ D {\bf 84} (2011) 054509
  [arXiv:1102.5300 [hep-lat]].

\bibitem{Bornyakov:2015eaa}
  V.G.~Bornyakov,  R.~Horsley, R.~Hudspith, Y.~Nakamura, H.~Perlt, D.~Pleiter, P.E.L.~Rakow, G.~Schierholz, A.~Schiller, H.~St\"uben and J.M.~Zanotti,
  arXiv:1508.05916 [hep-lat],
  PoS LATTICE {\bf 2015} (2016) 264 [arXiv:1512.05745 [hep-lat]].

\bibitem{cooling}
  B.~Berg,
  Phys.\ Lett.\ B {\bf 104}, 475 (1981);
  Y.~Iwasaki and T.~Yoshie,
  Phys.\ Lett.\ B {\bf 131}, 159 (1983);
  M.~Teper,
  Phys.\ Lett.\ B {\bf 162}, 357 (1985);
  E.~-M.~Ilgenfritz, M.L.~Laursen, G.~Schierholz, M.~M\"uller-Preussker and   H.~Schiller,
  Nucl.\ Phys.\ B {\bf 268}, 693 (1986).

\bibitem{Bali:2017qce}
  G.~Bali, S.~Collins and J.~Simeth,
  EPJ Web Conf.\  {\bf 175} (2018) 05028
  [arXiv:1710.06733 [hep-lat]].

\bibitem{Eichhorn:2012uv}
  A.~Eichhorn, H.~Gies and D.~Roscher,
  Phys.\ Rev.\ D {\bf 86} (2012) 125014
  [arXiv:1208.0014 [hep-ph]].

\bibitem{Weinberg}
  See, for example, S.~Weinberg, ``The quantum theory of fields. Vol. 2: Modern applications'', Cambridge University Press (1996).

\end{thebibliography}
\end{document}